\documentclass{PoS}

\usepackage{overpic}
\usepackage{axodraw}

\newcommand{\be}{\begin{equation}}
\newcommand{\ee}{\end{equation}}
\newcommand{\ba}{\begin{eqnarray}}
\newcommand{\ea}{\end{eqnarray}}

\title{Status of Strong ChPT}

\ShortTitle{Status of Strong ChPT}

\author{\speaker{Johan Bijnens}%
        \\
        Department of Theoretical Physics, Lund University,
        S\"olvegatan 14A, SE 22362 Lund, Sweden\\
        E-mail: \email{bijnens@thep.lu.se}}

\abstract{The present status of Chiral Perturbation Theory (ChPT) in the strong
mesonic sector is discussed. A very short introduction to ChPT is followed
by an overview of existing two-loop calculations and the determination
of the Low-Energy-Constants (LECs). I discuss the case of $\eta\to\pi\pi\pi$
decays in somewhat more detail and finish by mentioning some recent
work on partially quenched ChPT and on the renormalization group in ChPT.
}

\FullConference{International Workshop on Effective Field Theories: 
  from the pion to the upsilon\\
                 2-6 February 2009\\
                 Valencia, Spain}

\renewcommand{\logo}
{\raisebox{-5mm}
{\parbox{\textwidth}{\begin{flushright}
LU TP 09-09\\
April 2009
\end{flushright}
\includegraphics{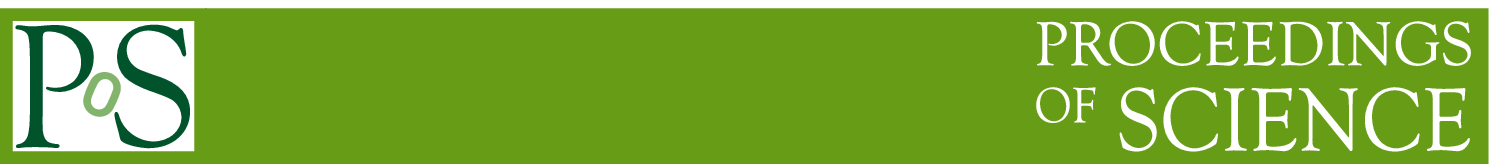}}
}}
\FullConference{{\rm To be published in the proceedings of:\\}
International Workshop on Effective Field Theories: 
  from the pion to the upsilon\\
                 2-6 February 2009\\
                 Valencia, Spain}

\begin{document}

 In this talk I will restrict myself to the light
pseudo-scalar mesons (two, three or more flavours) and strong interaction
only, including interaction with external currents and densities.
This excludes the work including baryons, heavy quarks, vector mesons,
structure functions and related quantities and non-leptonic
weak interactions as well as effects of photon loops. I start by mentioning a
few important historical papers which have roughly a jubilee this year,
Section \ref{history} followed by a short introduction
to Chiral Perturbation Theory (ChPT)
including discussions on chiral logarithms
and what exactly to expand in. The next two sections give
an overview of two and three flavour ChPT respectively,
where in the latter case
I go into some detail into how the LECs are determined experimentally
and the assumptions on the order $p^6$ LECs. Section \ref{sectioneta}
discusses the decay $\eta\to3\pi$ while
at the end I make some comments about recent work in
partially quenched ChPT and applications of the renormalization group
to ChPT. Many part are also in my earlier talk \cite{lattice07}.

\section{Some History: 50, 40, 35, 30, 25, 20 and 15 years ago}
\label{history}

Since in this conference we celebrate the scientific achievements of
Gerhard Ecker and J\"urg Gasser, it is appropriate to look back at the
history of Chiral Perturbation Theory.
I have picked out a few papers
which fell at or close to jubileum years. About 50 years ago the subject was
started with the Goldberger-Treiman relation \cite{GB} and
the advent of PCAC, the partially conserved axial-current \cite{PCAC},
and how this reproduced the Goldberger-Treiman relation.
About 40 years ago a lot of work had been done within the framework
of PCAC but 1968 and 1969 saw some very important papers:
the
Gell-Mann--Oakes--Renner relation \cite{GMOR} and the proper way how to
implement chiral symmetry in all generality in phenomenological
Lagrangians \cite{CCWZ}. Shortly afterwards loop calculations started
with e.g. loop results for $\pi\pi$ scattering \cite{EH}
and $\eta\to3\pi$ \cite{LP}. 30 years ago the start with the modern way of
including higher order Lagrangians and performing a consistent renormalization
came with \cite{Weinberg0}. At the same time there was also the beautiful
paper by Gasser and Zepeda about the types of non-analytical corrections that
can appear \cite{GZ}. The seminal papers by Gasser and Leutwyler of 25 years
ago then put the entire subject on a modern firm footing \cite{GL0,GL1}.
The same period also had my own entry into the subject \cite{BSW}.
Lots of one-loop calculations were done and the understanding that
the coefficients in the higher-order Lagrangians could be understood
from the contributions of resonances
was put on a firm footing \cite{GL0,EGPR}.
Let me close this historical part with two 15 year old papers,
a very clear discussion of the basics of ChPT \cite{Leutwyler1} and
the first full two-loop calculation \cite{BGS}.

\section{Chiral Perturbation Theory: ChPT, CHPT or $\chi$PT}

ChPT is best described as
``\emph{
Exploring the consequences of the chiral symmetry of QCD
and its spontaneous breaking using
effective field theory techniques}''
and a particularly clear discussion about its derivation and underlying
assumptions can be found in \cite{Leutwyler1}.
Some reviews are
\cite{review1,reviewp6}. More reviews and
references to introductory lectures can be found on the webpage
\cite{website}. 

For effective field theories, there are three principles that are needed
and for ChPT they are
\begin{itemize}
\parskip0cm\itemsep0cm
\item {\bf Degrees of freedom:} Goldstone Bosons from
the spontaneous chiral symmetry breakdown.
\item {\bf Power counting:} This is what allows a systematic ordering of terms
and is here essentially dimensional counting in momenta and masses.
\item {\bf Expected breakdown scale:} The scale of the not explicitly included
physics, here resonances, so the scale is of order $M_\rho$, 
but this is channel dependent. 
\end{itemize}

Chiral symmetry is the (continuous) interchange of quarks. If we look at the
QCD Lagrangian
\ba
{\cal L}_{QCD} &=&  \sum_{q=u,d,s}
\left[i \bar q_L D\hskip-1.3ex/\, q_L +i \bar q_R D\hskip-1.3ex/\, q_R
- m_q\left(\bar q_R q_L + \bar q_L q_R \right)
\right]-\frac{1}{4} G^a_{\mu\nu} G^{a\mu\nu}
\ea
we see that we have an $SU(3)_V$ symmetry for equal quark masses but
for $m_q=0$ we can change left- and right-handed separately giving a
$SU(3)_L\times SU(3)_R$ symmetry with
 $q^T = (\begin{array}{ccc}u & d& s\end{array})$ transforming as
$q_L\to g_Lq_L$ and $q_R\to g_R q_R$.

\begin{figure}[ht]
\vskip-2cm
\hskip1cm
\begin{minipage}{0.3\textwidth}
\unitlength=0.5pt
\begin{picture}(100,100)
\SetScale{0.5}
\SetWidth{1.5}
\Line(0,100)(100,0)
\Line(0,0)(100,100)
\Vertex(50,50){5}
\end{picture}
\hfill\raisebox{25pt}{$p^2$}\\[0.25cm]
\unitlength=0.5pt
\begin{picture}(100,30)
\SetScale{0.5}
\SetWidth{1.5}
\Line(0,15)(100,15)
\end{picture}
\hfill\raisebox{5pt}{$1/p^2$}\\[0.25cm]
$\int d^4p$\hfill$p^4$
\end{minipage}
\hfill
\raisebox{0.75cm}{
\begin{minipage}{0.45\textwidth}
\begin{picture}(100,100)
\SetScale{0.5}
\SetWidth{1.5}
\Line(0,100)(20,50)
\Line(0,0)(20,50)
\Vertex(20,50){5}
\CArc(50,50)(30,0,180)
\CArc(50,50)(30,180,360)
\Vertex(80,50){5}
\Line(80,50)(100,100)
\Line(80,50)(100,0)
\end{picture}
\hskip-1cm\raisebox{25pt}
{$(p^2)^2\,(1/p^2)^2\,p^4 = p^4$}\\[0.25cm]
\unitlength=0.5pt
\begin{picture}(100,100)
\SetScale{0.5}
\SetWidth{1.5}
\Line(0,0)(50,40)
\Line(0,50)(50,40)
\CArc(50,70)(30,0,180)
\CArc(50,70)(30,180,360)
\Vertex(50,40){5}
\Line(50,40)(100,50)
\Line(50,40)(100,0)
\end{picture}
~~\raisebox{25pt}
{$(p^2)\,(1/p^2)\,p^4 = p^4$}
\end{minipage}
}
\caption{\label{figpower} An illustration of the power-counting in ChPT.
On the left we have the lowest order vertex with two powers of momenta
or masses,
the meson propagator with two inverse powers and the loop integration
leading to four
powers. On the right hand-side we see two one-loop contributions and how the
counting on the left leads to the same power $p^4$ for both diagrams.
This counting can be generalized to all orders \cite{Weinberg0}.}
\end{figure}
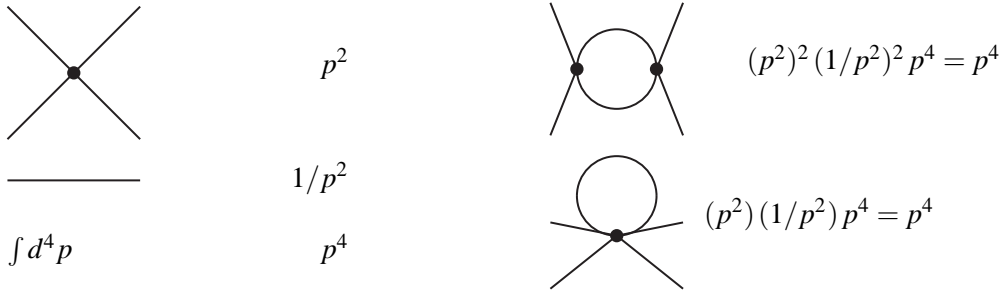

The chiral symmetry is broken spontaneously by vacuum condensates
$\langle \bar q q\rangle = \langle \bar q_L q_R+\bar q_R q_L\rangle
\ne 0$. This breaks the eight axial generators of the symmetry group but
leaves the vector part unbroken: $SU(3)_L\times SU(3)_R\to SU(3)_V$.
This produces eight massless Goldstone Bosons \emph{and} their
interaction vanishes at zero momentum. The latter is very important, it is
the reason why there exists a proper power-counting in ChPT.
This is illustrated in Fig.~\ref{figpower}.

We now start to look at the needed Lagrangians.
The $SU(3)_L\times SU(3)_R/SU(3)_V$ manifold is parameterized by a matrix
\be
U=\exp\left(i\sqrt(2)\Phi/F_0\right)
\quad\mbox{with}\quad
\Phi (x) = \,
{ \left( \begin{array}{ccc}
\frac{ \pi^0}{ \sqrt 2} \, + \, \frac{ \eta_8}{ \sqrt 6}
 & \pi^+ & K^+ \\
\pi^- & - \frac{\pi^0}{\sqrt 2} \, + \, \frac{ \eta_8}
{\sqrt 6}    & K^0 \\
K^- & \bar K^0 & - \frac{ 2 \, \eta_8}{\sqrt 6}
\end{array}  \right)} .
\ee
The traceless Hermitian matrix $\Phi$ is written in the usual
pseudo-scalar fields. With the covariant derivative
$
D_\mu U = \partial_\mu U -i r_\mu U + i U l_\mu \,,
$
which includes the
left and right external currents: $
r(l)_\mu = v_\mu +(-) a_\mu$ and the matrix $\chi = 2 B_0 (s+ip)$
that contains the external scalar and pseudo-scalar fields $s$ and $p$,
  the lowest order Lagrangian is
\be
\label{lagLO}
{\cal L}_2 = \left({F_0^2}/{4}\right)
\left\{\langle D_\mu U^\dagger D^\mu U \rangle 
+\langle \chi^\dagger U+\chi U^\dagger \rangle \right\}\, .
\ee
$\langle A \rangle$ is the trace over flavours $Tr_F\left(A\right)$.
Quark masses are included via $s= \mathrm{diag}(m_u,m_d,m_s) + \cdots$.

At higher orders the number of terms in the Lagrangian increases
rapidly. There exist two types,
those representing contact terms, i.e. without pseudo-scalar bosons,
and those with. The former can never be measured but are the reflection
in ChPT of the definition of currents and densities in QCD.
The latter are usually called low-energy-constants (LECs).
The number of parameters at the various orders
is shown in Tab.~\ref{tabparams}.
\begin{table}[t]
\begin{center}
\begin{tabular}{ccccccc}
      & \multicolumn{2}{c}{ 2 flavour} & \multicolumn{2}{c}{3 flavour} &
\multicolumn{2}{c}{ 3+3 PQChPT}\\
\hline
$p^2$ & $F,B$ & 2 & $F_0,B_0$ & 2 &  $F_0,B_0$ &  2 \\
$p^4$ & $l_i^r,h_i^r$ & 7+3 & $L_i^r,H_i^r$ & 10+2 & 
      $\hat L_i^r,\hat H_i^r$ &  11+2 \\
$p^6$ & $c_i^r$ & 52+4 & $C_i^r$ & 90+4 &  $K_i^r$ &
       112+3\\
\hline
\end{tabular}
\end{center}
\caption{\label{tabparams} The number of parameters+contact terms for the
various types of ChPT.}
\end{table}
The order $p^2$ is from \cite{Weinberg1}, order $p^4$ from \cite{GL0,GL1},
order $p^6$ from \cite{BCE1} after an earlier partial result \cite{FS2}.
The partially quenched results are derived from the $n_F$ flavour
case \cite{BDL1}. The difficulty in obtaining a minimal set can be seen
from the recent discovery of a new relation for two flavours \cite{Haefeli1}.
Since the normal case is a continuous limit of the partially quenched
case, the resulting LECs are linear combinations of
partially quenched LECs
using the Cayley-Hamilton relations given in \cite{BCE1}.
The general divergence structure at this order is known \cite{BCE2}.
The parameters $B\ne B_0$ and $F\ne F_0$ are the two versus three-flavour
lowest order constants, these are different quantities.

The main predictions of ChPT are twofold. 1) It
relates processes with different numbers of pseudo-scalars. 2) 
It predicts nonanalytic dependences at higher orders, often referred to
generically as \emph{Chiral Log(arithm)s}. As an example, the pion mass
for $n_F=2$ is given at NLO by \cite{GL0}
\be
\label{mpiNLO}
m_\pi^2 = 2 B \hat m  + \left(\frac{2 B \hat m}{F}\right)^2
\left[ \frac{1}{32\pi^2}{\log\frac{\left(2 B \hat m\right)}{\mu^2}} 
+ 2 l_3^r(\mu)\right] +\cdots
\ee
The implicit $\mu$ dependence in $l_3^r$
and the explicit dependence in the
logarithm cancel.

The LECs, like $l_3^r$ in (\ref{mpiNLO}),
have to be determined experimentally or from lattice calculations.
For $n_F=2$ Ref.~\cite{GL0} introduced the $\mu$ independent
$
\bar l_i = \left(32\pi^2/\gamma_i\right)\, l_i^r(\mu)
-\log\left(M_\pi^2/\mu^2\right)\,,
$
which are proportional to the LECs $l_i^r(\mu=m_\pi)$.
For $n_F=3$ some of the corresponding $\gamma_i$ are zero and no good
equivalent definition of $\bar L_i$ exists. Here we always quote the
$L_i^r(\mu)$. The scale $\mu$ is arbitrary but becomes relevant
when using estimates for higher order constants.

A question which is often misunderstood is what quantities to expand in.
The ChPT expansion is in momenta and masses. However, one first has to decide
whether to expand in lowest order quantities, like $F,2B\hat m$, or
physical masses and decay constants, like $m_\pi,m_K,m_\eta,F_\pi,F_K$.
The latter is not unique either since the
Gell-Mann--Okubo relation and kinematical relations like
$s+t+u=2m_\pi^2+2m_K^2$ for $\pi K$-scattering can be (and are heavily)
used to rewrite expressions. This sounds trivial but can change much how a
series convergence looks as shown below for a simple example.
I prefer to
use physical masses and decay constants rather than the lowest
order quantities. The physical quantities are typically better known
and the chiral logs are created by particles propagating with their physical
momentum. Also, thresholds appear in the right places at each order in
perturbation theory. The differences are higher order,
but can be numerically important.

\begin{figure}[ht]
\includegraphics[width=0.49\textwidth]{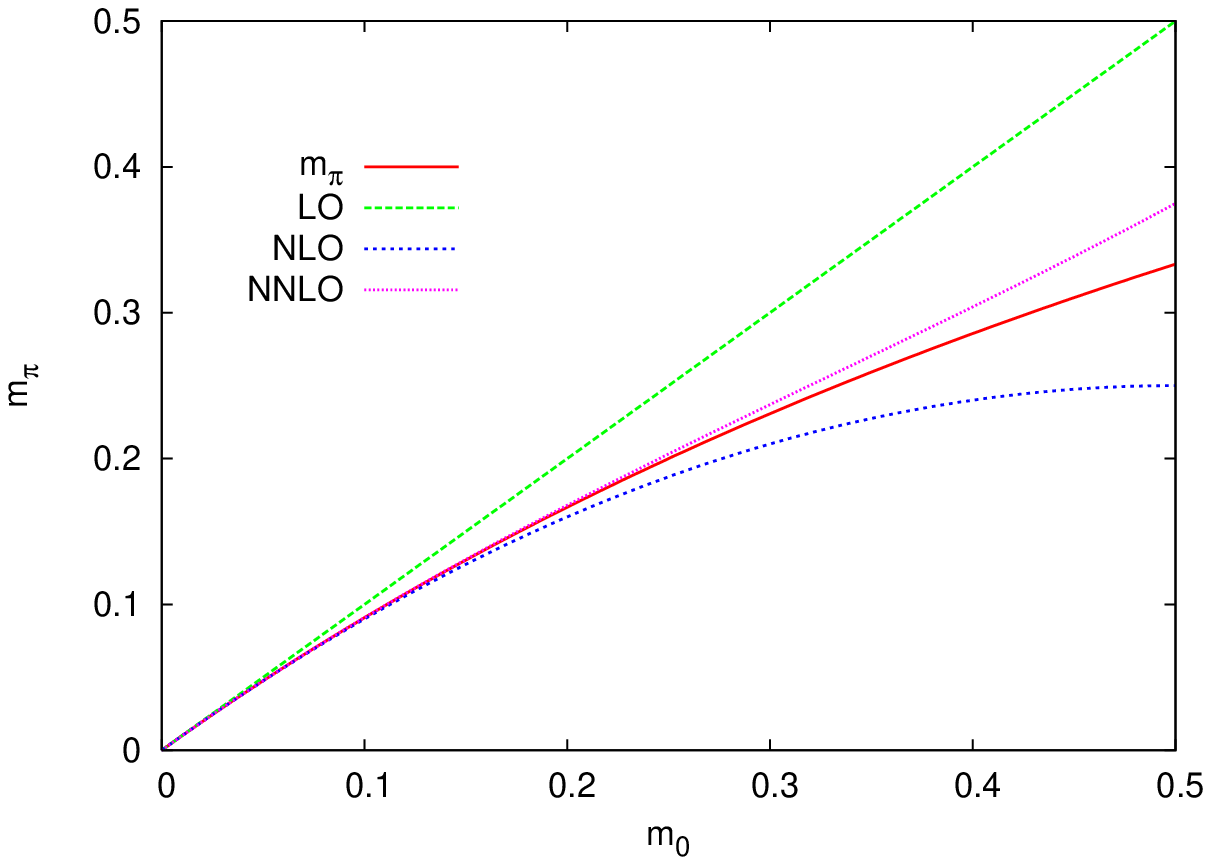}
\includegraphics[width=0.49\textwidth]{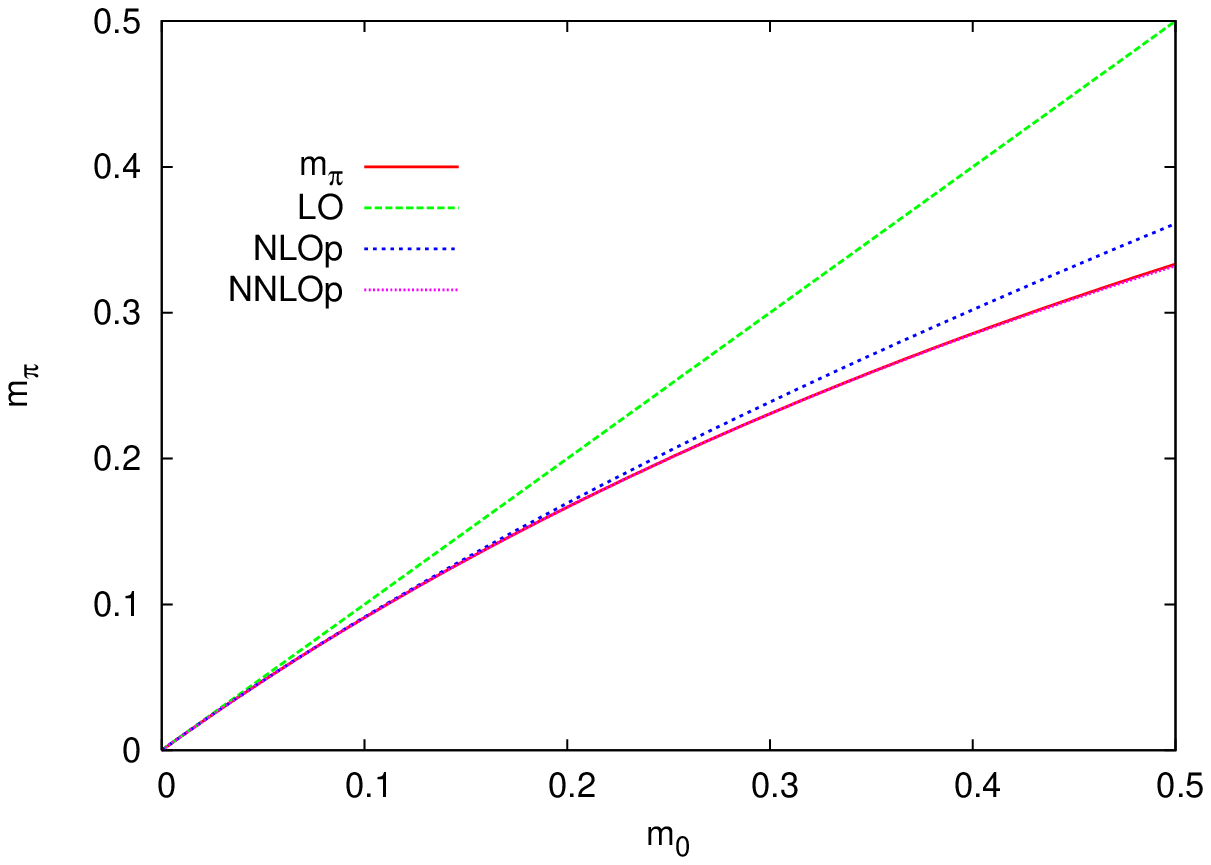}
\caption{On the left the expansion of $m_\pi$
in terms of $m_o/f_0$ 
(\protect\ref{mpi0}) 
and on the right 
in terms
of $m_\pi/f_\pi$ 
(\protect\ref{mpi1}). Shown are the full results ($m_\pi$) and the first three
approximations.}
\label{figexample} 
\end{figure}

 Take as a simple example the relations
$
m_\pi = m_0/\left(1+am_0/f_0\right),\quad
m_\pi = f_0/\left(1+bm_0/f_0\right)\,,
$
as exact. We can expand to NNLO in several ways, some examples are
\ba
\label{mpi0}
m_\pi = m_0 -a \frac{m_0^2}{f_0} + a^2 \frac{m_0^3}{f_0^2}+\cdots
&\quad&
f_\pi = f_0 \left(1 -b \frac{m_0}{f_0} + b^2 \frac{m_0^2}{f_0^2}+\cdots\right)
\\
\label{mpi1}
 m_\pi = m_0 -a \frac{m_\pi^2}{f_\pi} + a (b-a) \frac{m_\pi^3}{f_\pi^2}+\cdots
&&
f_\pi = f_0 \left(1 -b \frac{m_\pi}{f_\pi} + b (2b-a) \frac{m_\pi^2}{f_\pi^2}+\cdots\right)
\ea
The coefficients in the expansion and the actual numerical
values clearly depend on the way we write the results.
The plots in Fig.~\ref{figexample} show the convergence
for $a= 1$, $b= 0.5$ and $f_0=1$.
Only knowing the first three terms one would draw
very different conclusions on the quality of the convergence from
Fig.~\ref{figexample} for the different ways of expanding.

\section{Two-flavour ChPT at NNLO}

References to order $p^2$ and $p^4$ work can be found in \cite{reviewp6}.
The first work at NNLO used dispersive methods to obtain the nonanalytic
dependence on kinematical quantities, $q^2,s,t,u$ at NNLO. This was done
for the vector (electromagnetic) and scalar form-factor of the pion
in \cite{GM} (numerically) and \cite{CFU} (analytically) and for
$\pi\pi$-scattering analytically in \cite{Knechtpipi}. The work of
\cite{Knechtpipi} allowed to put many full NNLO calculations
in two-flavour ChPT in a simple analytical form.

Essentially all processes of interest are calculated to NNLO fully in ChPT
starting with $\gamma\gamma\to\pi^0\pi^0$ \cite{BGS,GIS1},
$\gamma\gamma\to\pi^+\pi^-$ \cite{Burgi1,GIS2},
$F_\pi$ and $m_\pi$ \cite{Burgi1,BCEGS1,BCT}, $\pi\pi$-scattering
\cite{BCEGS1}, the pion scalar and vector form-factors \cite{BCT}
and pion radiative decay $\pi\to\ell\nu\gamma$ \cite{BT1}.
The pion mass is known at order $p^6$ in finite volume \cite{CH}.
Recently $\pi^0\to\gamma\gamma$ has been done to this order
as discussed in the talk by Kampf \cite{Kampf1}.

The LECs have been fitted in several processes. $\bar l_4$ from fitting
to the pion scalar radius \cite{BT1,CGL}, $\bar l_3$ from an estimate of
the pion mass dependence on the quark masses \cite{GL0,CGL}
and $\bar l_1$, $\bar l_2$ from the agreement with 
$\pi\pi$-scattering \cite{CGL}, $\bar l_6$ from the pion charge radius 
\cite{BCT}
and $\bar l_6-\bar l_5$ from the axial form-factor in $\pi\to\ell\nu\gamma$.
There is also a recent determination of $\bar l_5$ from hadronic
tau decays \cite{PP}.
The final best values are \cite{BCT,BT1,CGL,PP}
\be
\label{valueli}
\begin{array}{llll}
\bar l_1=-0.4\pm 0.6\,,\quad&
\bar l_2 =4.3\pm0.1\,,&
\bar l_3=2.9\pm2.4\,,&
\bar l_4=4.4\pm0.2\,,
\\
\bar l_6-\bar l_5 = 3.0\pm0.3\,,\quad&
\bar l_6 = 16.0\pm0.5\pm0.7\,,\quad&
\bar l_5 = 12.24\pm0.21\,.&
\end{array}
\ee

There is information on some combinations of $p^6$ LECs.
These are basically via the curvature in the vector and scalar form-factor
of the pion \cite{BCT} and two combinations from $\pi\pi$-scattering
\cite{CGL} from the knowledge of $b_5$ and $b_6$ in that reference.
The order $p^6$ LECs $c_i^r$ are estimated 
to have a small effect for
$m_\pi,f_\pi$ and $\pi\pi$-scattering.

\begin{figure}
\includegraphics[width=0.49\textwidth]{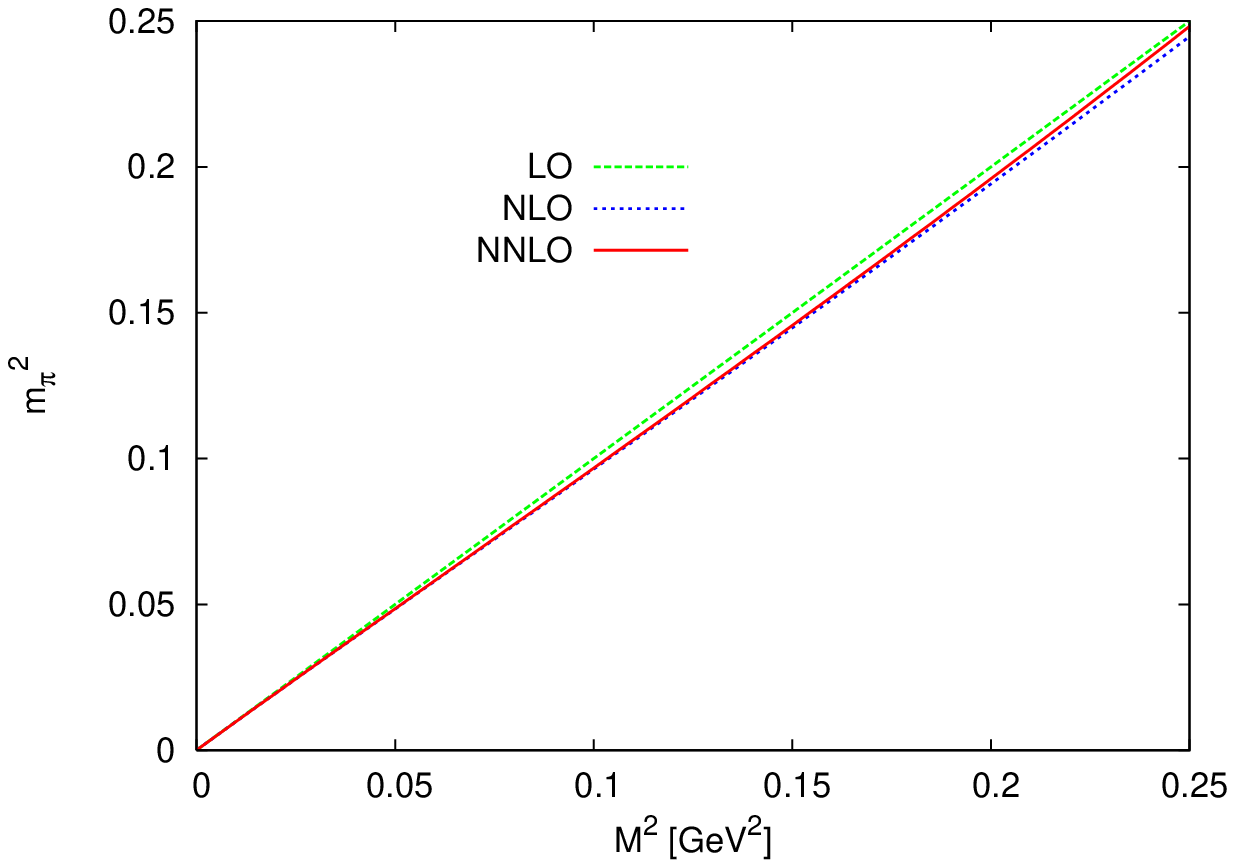}
\includegraphics[width=0.49\textwidth]{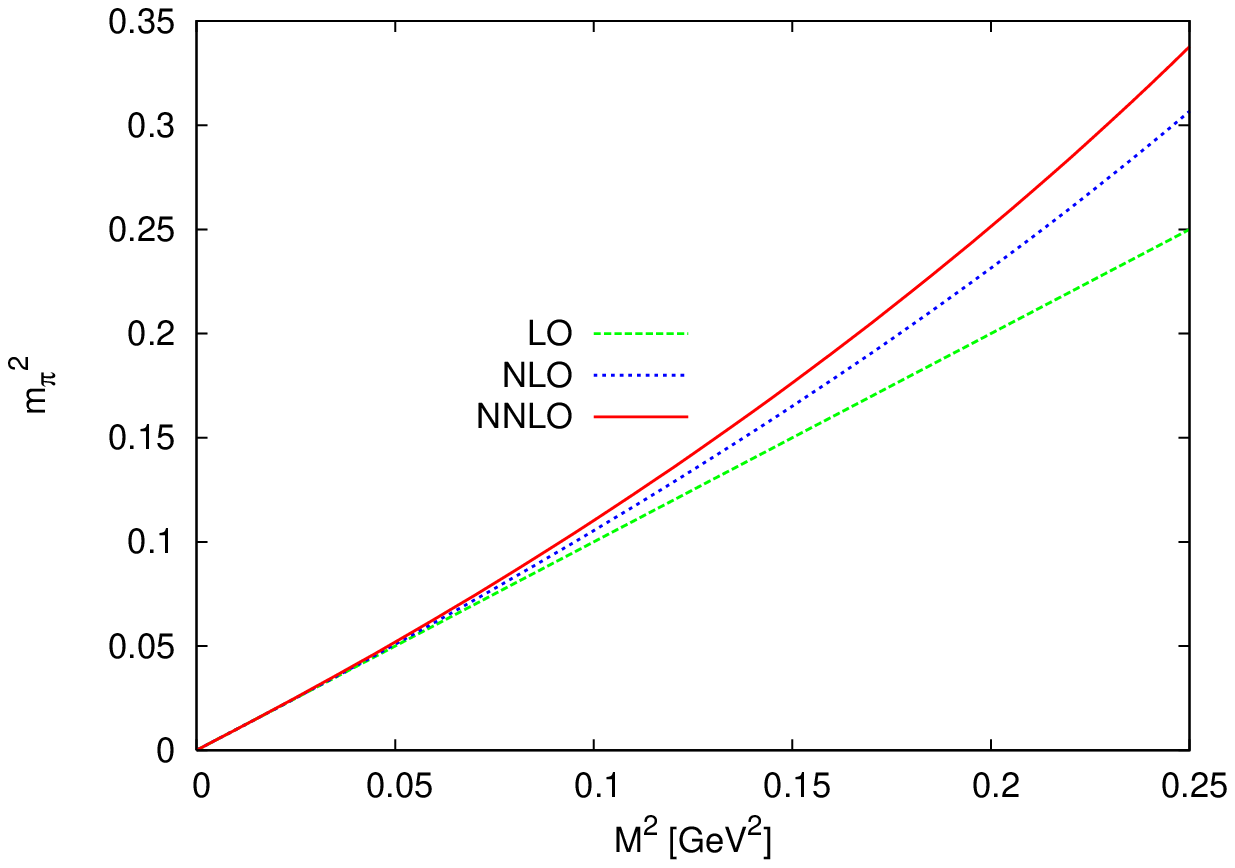}
\caption{The pion mass squared as a function of the quark mass via
$M^2=2B\hat m$, left with inputs as in (\protect\ref{valueli}) and right with
$\bar l_3=0$, both are for $n_F=2$ ChPT.}
\label{figmpi}
\end{figure}

\begin{figure}
\begin{minipage}{0.53\textwidth}
\includegraphics[width=\textwidth]{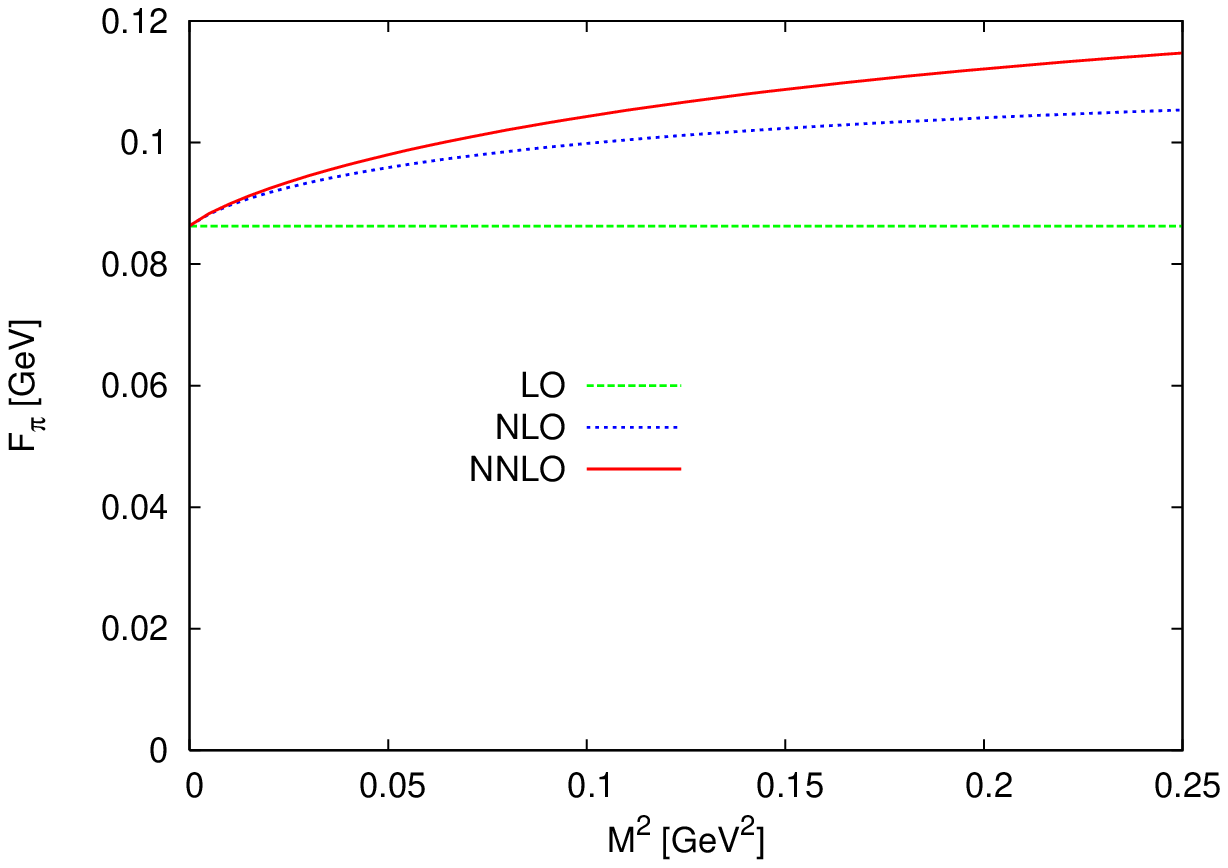}
\caption{The pion decay constant as a function of the quark mass via
$M^2=2B\hat m$, for $n_F=2$ ChPT.}
\label{figfpi}
\end{minipage}
~~
\begin{minipage}{0.45\textwidth}
\vskip-0.5cm
\includegraphics[angle=-90,width=\textwidth]{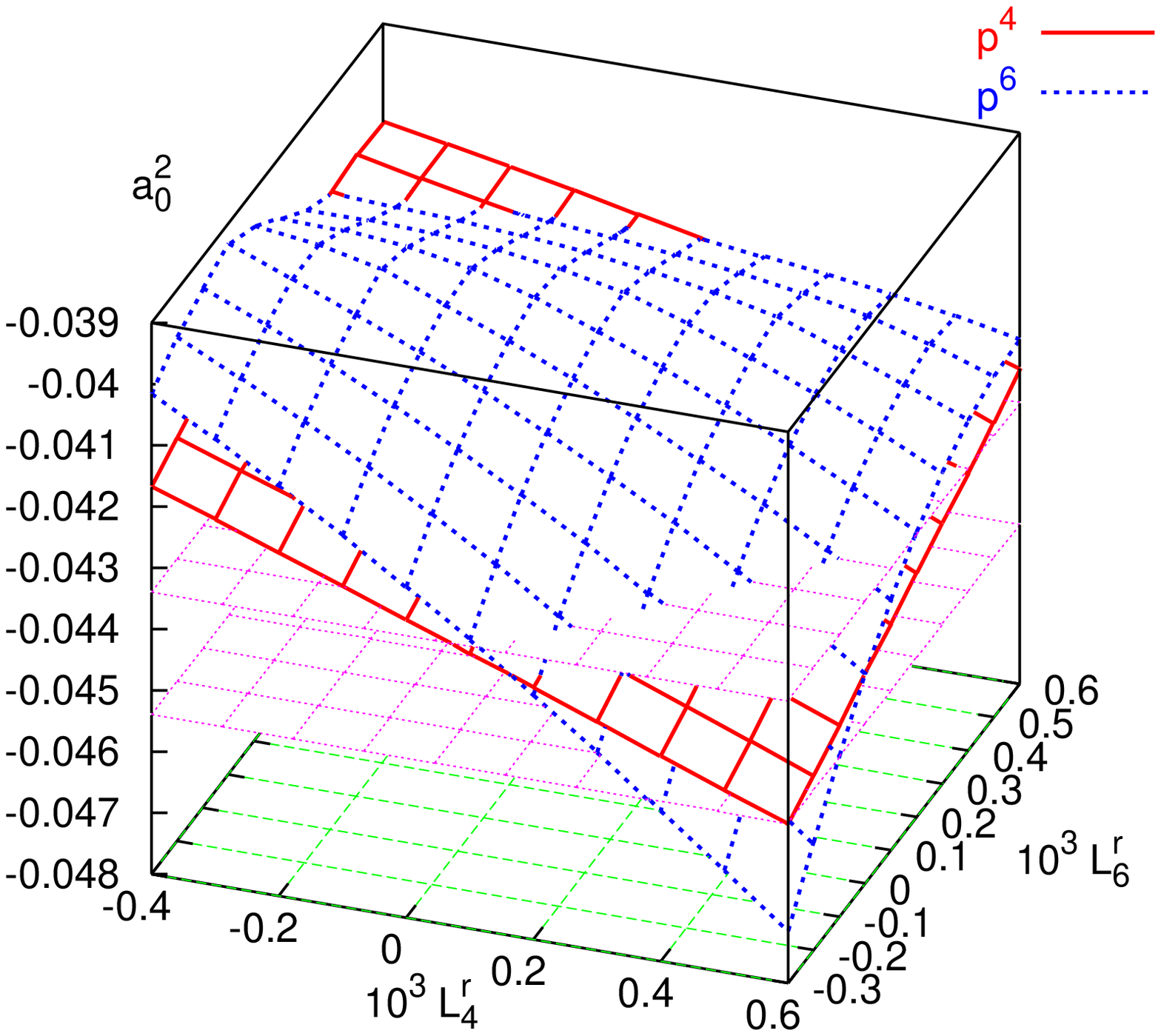}
\caption{The $\pi\pi$ scattering length $a^2_0$ in three-flavour ChPT as
a function of the input values of $L_4^r,L_6^r$ used in the fits,
 from \cite{BDT1}.}
\label{figpipi}
\end{minipage}
\end{figure}

Let me now show the dependence on the quark mass via $M^2 = 2B \hat m$
for $m_\pi^2$ with surprisingly  small NLO and NNLO corrections
for the values of 
the input parameters in (\ref{valueli}) and $c_i^r(\mu=0.77~\mathrm{GeV})=0$.
The full result is extremely linear as can be seen in the left plot in
Fig.~\ref{figmpi}.
The linearity is a consequence of the fitting parameters as can be seen in
the right figure in Fig.~\ref{figmpi}. Similarly, $F_\pi$ as a function
of $M^2$ expanded as in (\ref{mpi1}) is shown in Fig.~\ref{figfpi}.
The values of $m_\pi^2$, $F_\pi$ and $M^2$ are determined selfconsistently
via an iterative method from the ChPT formulas quoted in \cite{BCT}.

\section{Three-flavour ChPT}

\subsection{Calculations}

In this section I  discuss several results at NNLO in mesonic three-flavour
ChPT. The formulas here are much more involved than in
two-flavour ChPT and while the expressions have been reduced to a series
of well-defined two-loop integrals, the latter are evaluated numerically.
Both are the consequence of the different masses present.
The vector two-point functions
\cite{ABT1,GK1} and the isospin breaking in the $\rho\omega$
channel \cite{Maltman} were among the first calculated.
The disconnected scalar two-point function
relevant for bounds on $L_4^r$ and $L_6^r$ was worked out in \cite{Moussallam}
The remaining scalar two-point functions are known but unpublished
\cite{Bijnensscalar}. Masses and decay constants as well as axial-vector
two-point functions were the first calculations which required full two-loop
integrals, done in the $\pi$ and $\eta$ \cite{ABT1,GK2} and the $K$ channel
\cite{ABT1}. Including isospin breaking contributions to masses and decay
constants was done in \cite{ABT4}.
After $K_{\ell4}$ had also been
evaluated to NNLO \cite{ABT3} a fit to the LECs was done as described
below. The vacuum expectation values
in the isospin limit were done in \cite{ABT3},
with isospin breaking in \cite{ABT4} and at finite volume in~\cite{BG1}.

Vector (electromagnetic) form-factors for pions and kaons were calculated
in \cite{PS1,BT2} and in \cite{BT2} a NNLO fit for $L_9^r$ was performed.
$L_{10}^r$ can be had from hadronic tau decays \cite{PP}
or the axial form-factor in $\pi,K\to\ell\nu\gamma$.
The NNLO calculation is done, but no data fitting was performed\cite{Geng}.
A rather important calculation is the $K_{\ell3}$ form-factor. This calculation
was done by \cite{BT3,PS3} and a rather interesting relation between
the value at zero, the slope and the curvature for the scalar form-factor
obtained \cite{BT3}. Isospin-breaking has been included as well \cite{BG3}.

Scalar form-factors including sigma terms and scalar radii
\cite{BD} and $\pi\pi$ \cite{BDT1} and $\pi K$-scattering \cite{BDT2}
have been performed as well and used to place limits on $L_4^r$ and $L_6^r$.
Finally, the relations between the $l_i^r,c_i^r$ and $L_i^r,C_i^r$
 have been extended
to the accuracy needed to compare order $p^6$ results in two and three-flavour
calculations \cite{Haefeli2} and there has been some progress towards fully
analytical results for
$m_\pi^2$ \cite{Kaiser} and $\pi K$-scattering lengths \cite{KS}.
The most recent results are
 $\eta\to3\pi$ \cite{BG2}, isospin breaking in $K_{\ell3}$ \cite{BG3}.

\subsection{The fitting and results}

The inputs used for the fitting, as discussed more extensively in
\cite{ABT4,ABT3}, are
\begin{itemize}
\parskip0cm\itemsep0cm
\item $K_{\ell4}$: $F(0)$, $G(0)$, $\lambda$ from E865 at 
BNL\cite{Pislak1}.
\item $m^2_{\pi^0}$, $m^2_\eta$, $m_{K^+}^2$, $m_{K^0}^2$, electromagnetic
corrections include the violation of Dashen's theorem.
\item $F_{\pi^+}$ and $F_{K^+}/F_{\pi^+}$.
\item
$m_s/\hat m = 24$. Variations with
$m_s/\hat m$ were studied in \cite{ABT4,ABT3}.
\item
$L_4^r, L_6^r$ the main fit, 10, has them equal to zero, but see below
and the arguments in \cite{Moussallam}.
\end{itemize}
Some results of this fit are given in Tab.~\ref{tabfits}.
The errors are very correlated, see Fig.~6 in \cite{ABT3} for an example.
Varying the values of $L_4^r,L_6^r$ as input can be done with a
reasonable fitting chi-squared when varying $10^3 L_4^r$ from $-0.4$ to $0.6$
and $L_6^r$ from $-0.3$ to $0.6$ \cite{BD}.
The variation of many quantities with $L_4^r,L_6^r$
 (including the changes via the changed values of the other $L_i^r$) are shown
in \cite{BD,BDT1,BDT2}. Fit B was one of the fits with a good fit to the
pion scalar radius and fairly small corrections to the sigma terms \cite{BD}
while fit D \cite{Kazimierz} is the one that gave agreement with
$\pi\pi$ and $\pi K$-scattering threshold quantities.

\begin{table}[ht]
\begin{center}
\small
\begin{tabular}{ccccc}
                & fit 10 & same $p^4$ & fit B & fit D\\
\hline
$10^3 L_1^r$ & $0.43\pm0.12$ & $0.38$ & $0.44$ & $0.44$\\
$10^3 L_2^r$ & $0.73\pm0.12$ & $1.59$ & $0.60$ & $0.69$\\
$10^3 L_3^r$ & $-2.53\pm0.37$ & $-2.91$ &$-2.31$&$-2.33$\\
$10^3 L_4^r$ & $\equiv0$    & $\equiv 0$& $\equiv0.5$ & $\equiv0.2$\\
$10^3 L_5^r$ & $0.97\pm0.11$& $1.46$ & $0.82$ & $0.88$\\
$10^3 L_6^r$ & $\equiv0$    & $\equiv 0$& $\equiv0.1$ & $\equiv0$\\
$10^3 L_7^r$ & $-0.31\pm0.14$&$-0.49$ & $-0.26$ & $-0.28$\\
$10^3 L_8^r$ & $0.60\pm0.18$ & $1.00$ & $0.50$ & $0.54$\\
$10^3 L_9^r$ & $5.93\pm0.43$ & $7.0$  & --      &  -- \\
\hline
$2 B_0 \hat m/m_\pi^2$ & 0.736 & 0.991 & 1.129 & 0.958\\
$m_\pi^2$: $p^4,p^6$    & 0.006,0.258 & 0.009,$\equiv0$ & $-$0.138,0.009 &
          $-$0.091,0.133\\
$m_K^2$: $p^4,p^6$    & 0.007,0.306 & 0.075,$\equiv0$ & $-$0.149,0.094 &
          $-$0.096,0.201\\
$m_\eta^2$: $p^4,p^6$    & $-$0.052,0.318 & 0.013,$\equiv0$ & $-$0.197,0.073 &
          $-$0.151,0.197\\
$m_u/m_d$    & 0.45$\pm$0.05 & 0.52 & 0.52 & 0.50\\
\hline
$F_0$ [MeV]          & 87.7 & 81.1 & 70.4 & 80.4 \\
$\frac{F_K}{F_\pi}$: $p^4,p^6$ & 0.169,0.051 & 0.22,$\equiv0$ & 0.153,0.067 &
   0.159,0.061
\end{tabular}
\end{center}
\normalsize
\caption{The fits of the $L_i^r$ and some results, see text for
a detailed description. They are all quoted at $\mu=0.77$~GeV.
Table with values from \protect\cite{ABT4,BT2,BD,BDT2,Kazimierz}.}
\label{tabfits}
\end{table}

Note that $m_u/m_d=0$ is never even
close to the best fit and this remains true for the entire variation
with $L_4^r,L_6^r$. The value of $F_0$,
the pion decay constant in the three-flavour chiral limit,
can vary significantly, even though I believe that fit B is an extreme case.

\begin{figure}[ht]
\vskip-0.2cm
\begin{center}
\unitlength=0.4pt
\begin{overpic}[angle=-90,origin=rb,width=0.4\textwidth]{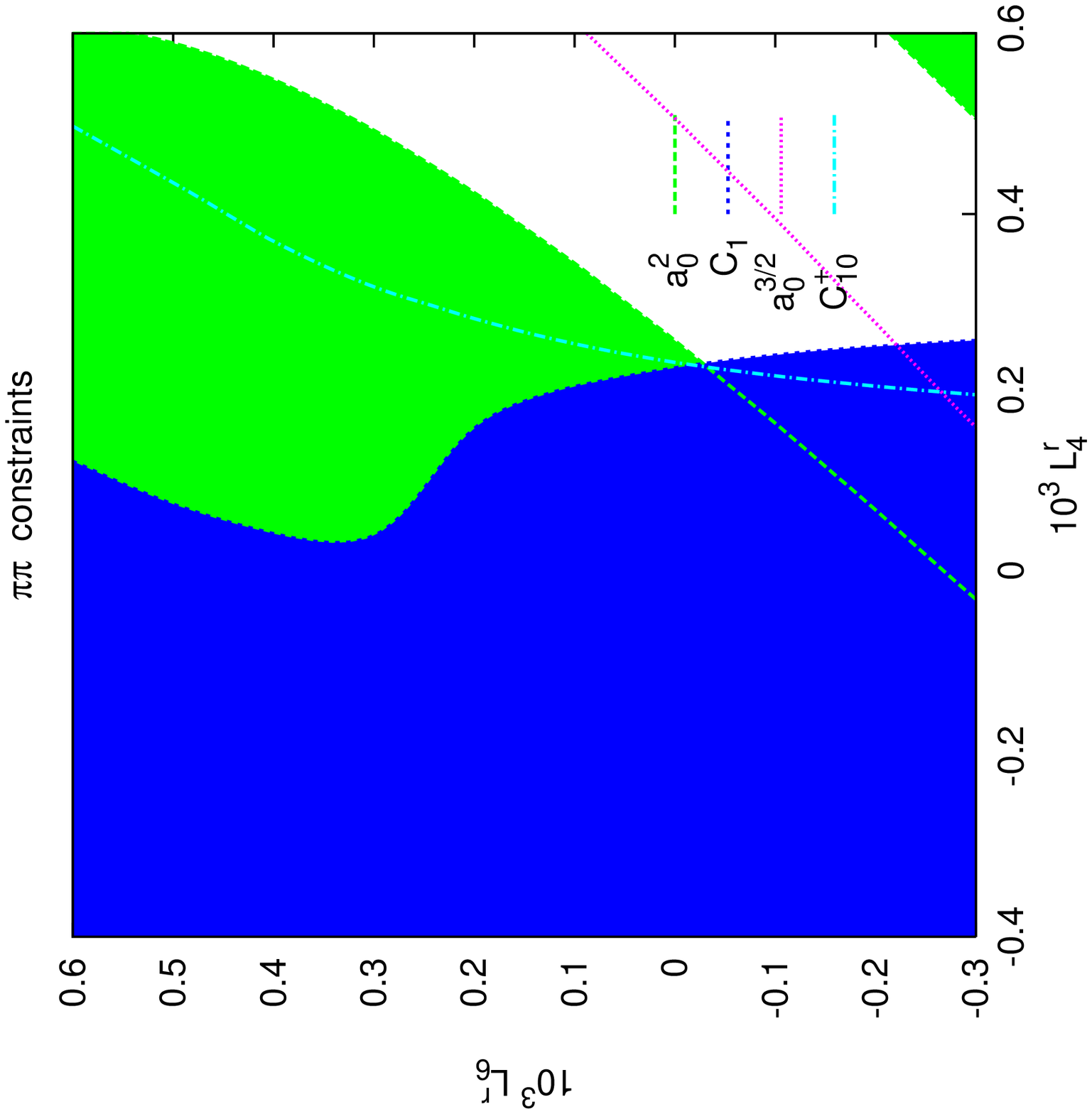}
\SetScale{0.4}
\SetWidth{1.5}
\CArc(285,155)(20,0,360)
\end{overpic}
\unitlength=0.4pt
\begin{overpic}[angle=-90,width=0.4\textwidth]{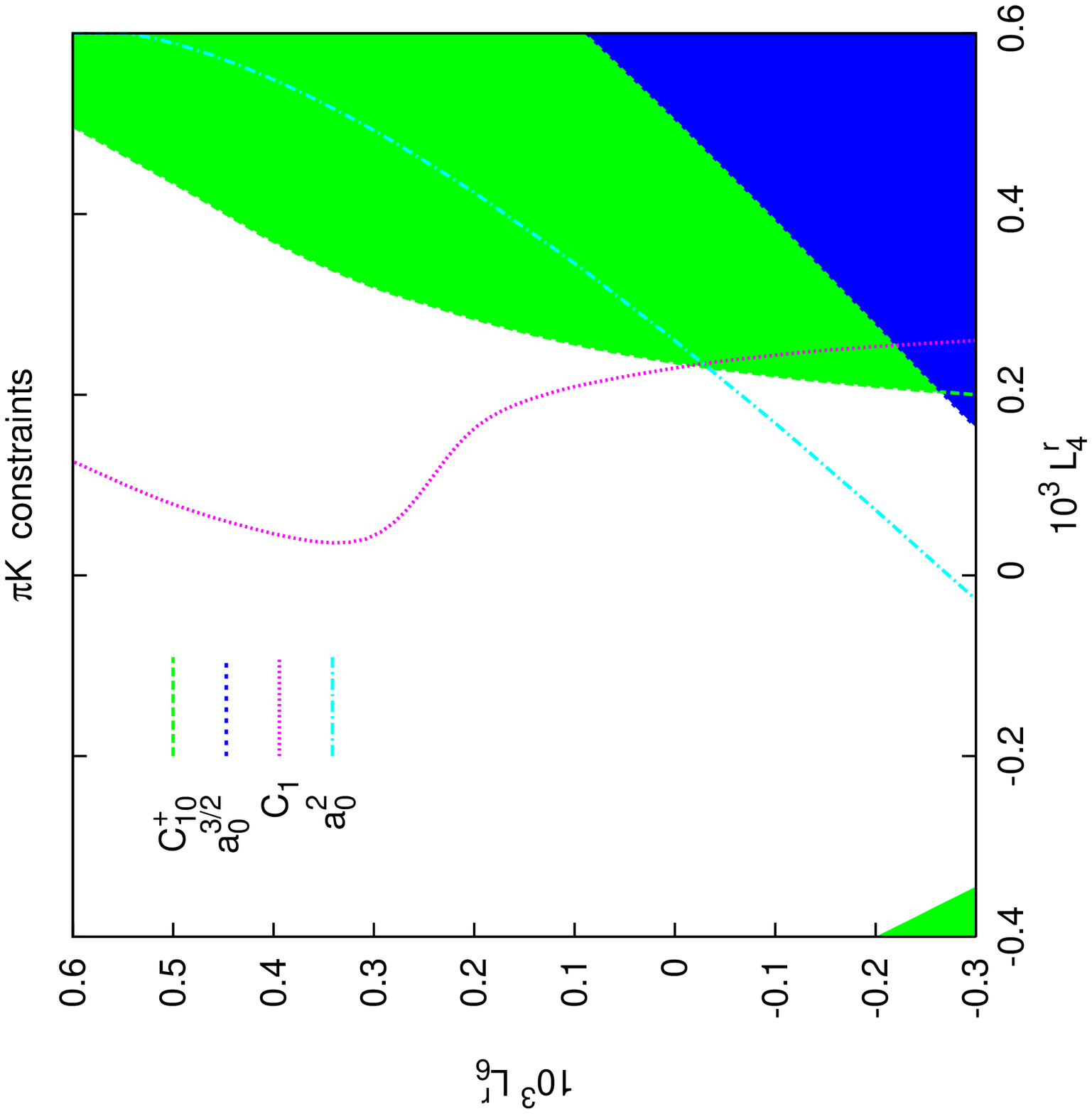}
\SetScale{0.4}
\SetWidth{1.5}
\CArc(275,160)(20,0,360)
\end{overpic}
\end{center}
\caption{The bounds on $L_4^r,L_6^r$ from $\pi\pi$ and $\pi K$-scattering
threshold parameters. Left $\pi\pi$ where the bound from $a^2_0$
shown in Fig.~\protect\ref{figpipi} is the most stringent.
Right $\pi K$. 
White regions are allowed.
The region of fit D, compatible with both, is indicated
by the circle. From \protect\cite{BDT2}.}
\label{figL4L6}
\end{figure}

In Fig.~\ref{figpipi} I show how the threshold parameter $a^2_0$
depend on the variation with  $L_4^r,L_6^r$. $a_0^0$ always agrees
well with the result of \cite{CGL} while $a^2_0$ only agrees well within a
limited region \cite{BDT1}. For comparison, the order $p^2$ values are
$a_0^0=0.159$ and $a^2_0=-0.0454$. The planes in Fig.~\ref{figpipi} indicate
the results $a_0^0=0.220\pm0.005$, $a^2_0= -0.0444\pm0.0010$ \cite{CGL}.
The same study was performed for $\pi K$ scattering lengths in \cite{BDT2}
with the results of the Roy-Steiner analysis \cite{BDM}. The resulting
limits on the input values of  $L_4^r,L_6^r$ are shown in Fig.~\ref{figL4L6}.
The resulting region called fit D in Tab.~\ref{tabfits} is
 $10^3 L_4^r \approx0.2$, $10^3 L_6^r \approx0.0$.
This general fitting obviously needs more work and systematic studies
and constraints from lattice QCD on $L_4^r,L_6^r$ will be very useful.

\begin{figure}
\vskip-0.5cm
\includegraphics[width=0.49\textwidth]{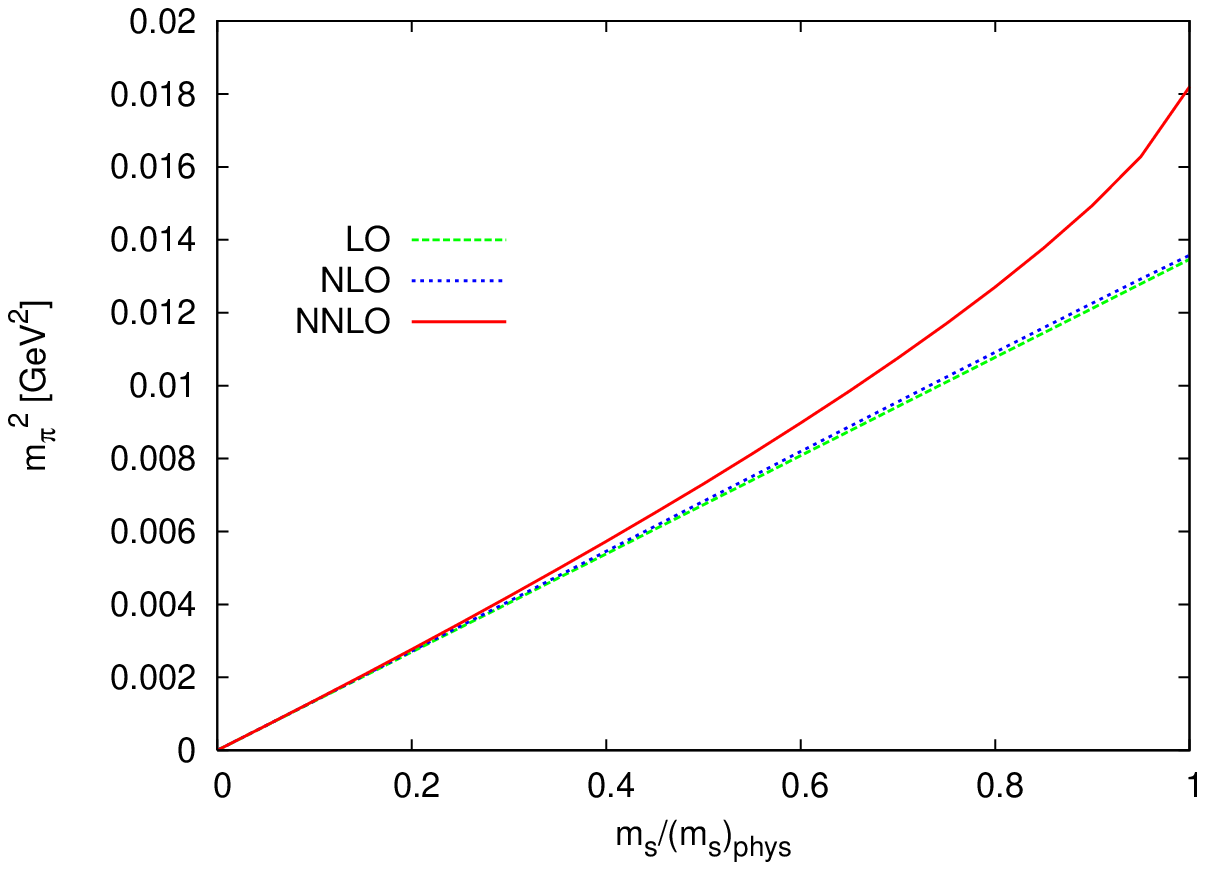}
\includegraphics[width=0.49\textwidth]{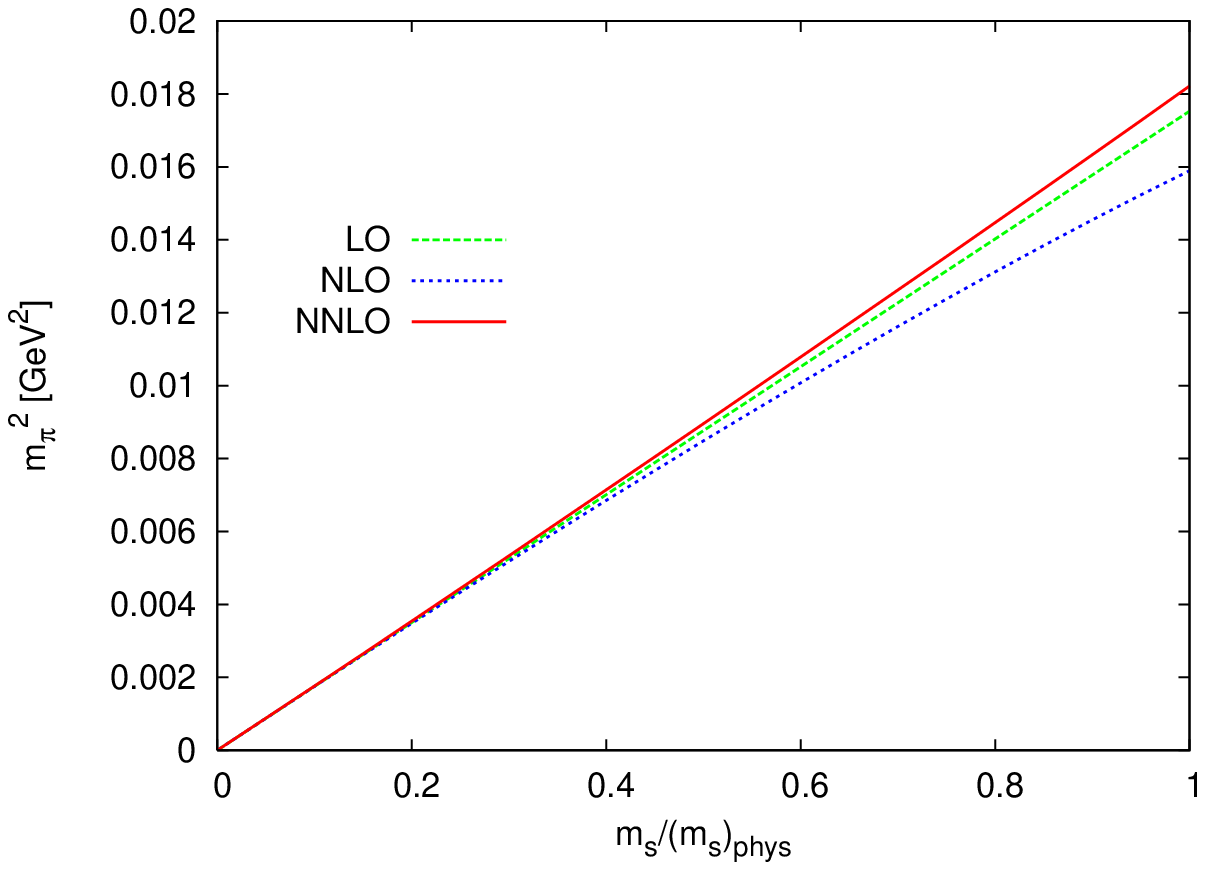}
\caption{$m_\pi^2$ as a function of $m_s$ for fit 10 
(left) and fit D (right) of 
Tab.~\protect\ref{tabfits} with $m_s/\hat m$ fixed. 
Note the difference in convergence properties between the two fits.}
\label{figmpinf3}
\end{figure}

I now show the dependence of a few quantities on the input masses.
These are updates of the plots shown in \cite{ABT3},
more can be found in \cite{lattice07}.
A selfconsistent set of $m_\pi^2$, $m_K^2$, $m_\eta^2$, $F_\pi$, $B_0 m_s$
and $B_0 \hat m$ with the fitted values of $L_i^r$ and $F_0$ is determined for
each input value of two masses. This is done by iterating the formulas
till convergence is reached.
I show $m_\pi^2$ for fit 10 and fit D keeping $m_s/\hat m=24$ and varying
$m_s$ in Fig.~\ref{figmpinf3}. 
The large corrections for fit 10 come from the kaon mass.
The decay constants ratio $F_K/F_\pi$ is shown as a function of $m_s$ 
with $m_s/\hat m = 24$ as well.
\begin{figure}
\begin{minipage}{0.44\textwidth}
\includegraphics[width=\textwidth]{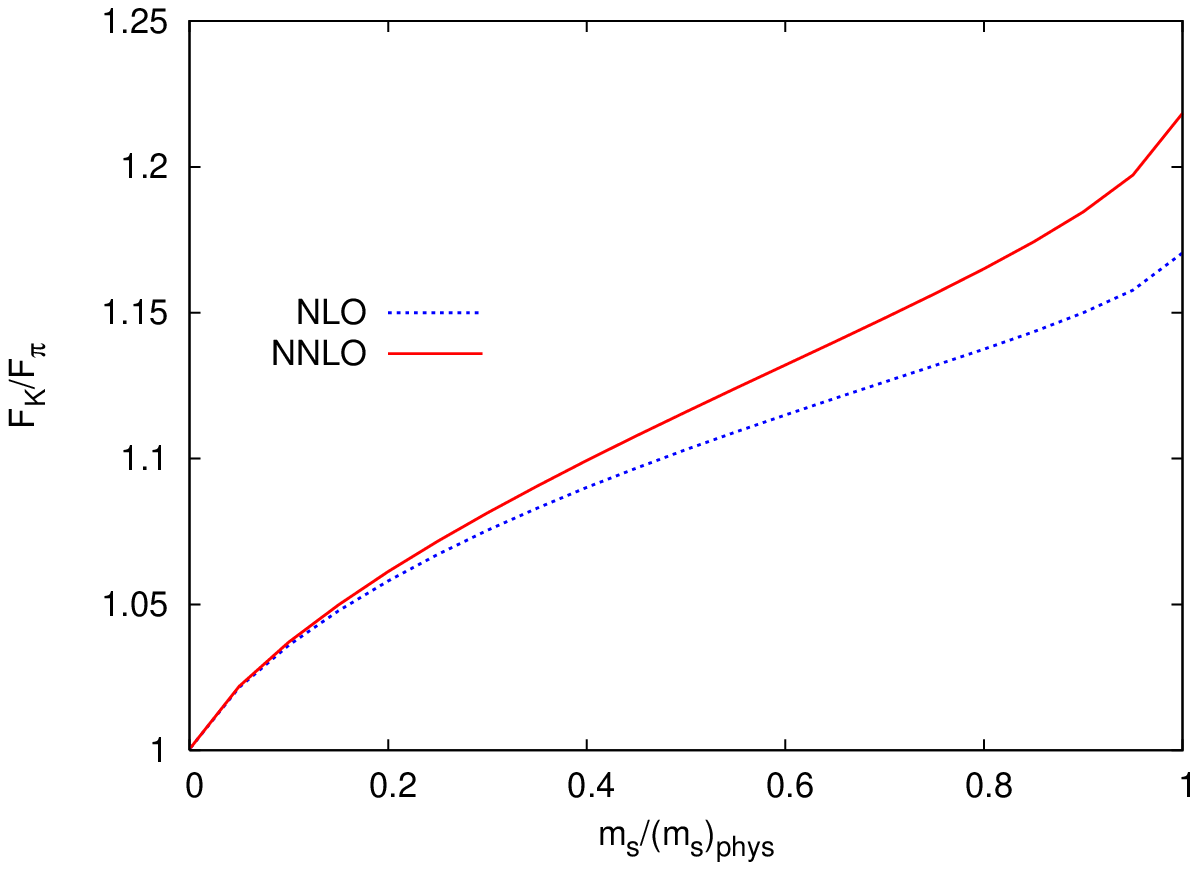}
\caption{For fit 10:  $F_K/F_\pi$ as a function of $m_s$
 with $m_s/\hat m$ fixed.}
\label{figfkmq}
\end{minipage}
~
\begin{minipage}{0.54\textwidth}
\rule{0cm}{2cm}\\
\unitlength=0.5pt
\begin{picture}(440,100)
\SetScale{0.5}
\SetWidth{1.5}
\Line(0,100)(10,50)
\Line(0,0)(10,50)
\Text(10,10)[]{$\pi$}
\Text(10,90)[]{$\pi$}
\Vertex(10,50){5}
\Line(10,52)(100,52)
\Line(10,48)(100,48)
\Text(55,65)[]{$\rho,S$}
\Text(55,30)[]{\small$\rightarrow q^2$}
\Vertex(100,50){5}
\Line(100,50)(110,0)
\Line(100,50)(110,100)
\Text(120,10)[]{$\pi$}
\Text(120,90)[]{$\pi$}
\Text(220,75)[]{\tiny$|q^2|<< m_\rho^2,m_S^2$}
\Text(220,50)[]{\Large$\Longrightarrow$}
\Line(320,100)(380,50)
\Line(320,0)(380,50)
\Line(380,50)(440,100)
\Line(380,50)(440,0)
\Text(380,75)[]{\tiny$C_i^r$}
\Vertex(380,50){5}
\end{picture}
\\
\rule{0cm}{0.8cm}
\caption{A schematic indication of the estimate of the order $p^6$ LECs by resonance exchange.}
\label{figCi}
\end{minipage}
\end{figure}

\subsection{$C_i^r$: estimates of order $p^6$ LECs}

Most numerical analysis at order $p^6$ use a (single) resonance
approximation to the order $p^6$ LECs. This is schematically shown in
Fig.~\ref{figCi}.
The main underlying motivation is the large $N_c$ limit and phenomenological
success at order $p^4$ \cite{EGPR}.
There is a large volume of work on this,
some references are \cite{MHA1}.
The numerical work I will report has used a rather simple resonance Lagrangian
\cite{EGPR,BCEGS1,ABT4,ABT3,EGPR}
only.
The estimates of the $C_i^r$ is the weakest point in the numerical
fitting at present, however, many results are not very sensitive to this.
The main problem is that the $C_i^r$ which contribute to the masses,
are estimated to be zero except for $\eta'$ effects and how these
might
affect the determination of the others. 
The estimate is $\mu$-independent
while the $C_i^r$ are not.

The fits done here in \cite{ABT4,ABT3,BD} try to check this by varying
the total resonance contribution by a factor of two, varying the scale $\mu$
from $550$ to $1000$~MeV and compare estimated $C_i^r$ to experimentally
determined ones. The latter works well, but the experimentally 
well determined ones are those with dependence on kinematic variables only,
not ones relevant for quark-mass dependence.

We are at present \cite{Jemos} working on a new fit and trying to find how
can be done without these estimates. That there might be some strain can be
seen from the different $C_i$ estimates from \cite{KM} shown in
Table \ref{tab:Ci} using the results of $\pi\pi$ and $\pi K$ scattering
of \cite{BDT1,BDT2}.

\begin{table}
\centerline{
\begin{tabular}{|c|cccc|}\hline
input & $C_1^r+4C_3^r$ & $C_2^r$      &$C_4^r+3C_3^r$&$C_1^r+4C_3^r+2C_2^r$ \\
\hline
$\pi K: C^+_{30}, C^+_{11}, C^-_{20}$
      & $20.7\pm 4.9$  & $-9.2\pm 4.9$ & $9.9\pm 2.5$&     $ 2.3\pm 10.8$   \\
$\pi K: C^+_{30}, C^+_{11}, C^-_{01}$
      & $28.1\pm 4.9$  & $-7.4\pm 4.9$ & $21.0\pm 2.5$&    $13.4\pm 10.8$ \\
$\pi\pi$
      & \              &               & $23.5\pm 2.3$&    $18.8\pm 7.2$  \\
 Resonance model
      & $7.2$           & $-0.5$        & $10.0$      &    $6.2$      \\ \hline
\end{tabular}}
\caption{ Different determinations of the same combinations
of the $C_i^r$ from $\pi\pi$ and $\pi K$ scattering \cite{KM}.}
\label{tab:Ci}
\end{table}

\section{$\eta\to \pi\pi\pi$}
\label{sectioneta}

\begin{figure}[h]
\vskip-0.5cm
\begin{minipage}{0.42\textwidth}
\includegraphics[width=\textwidth]{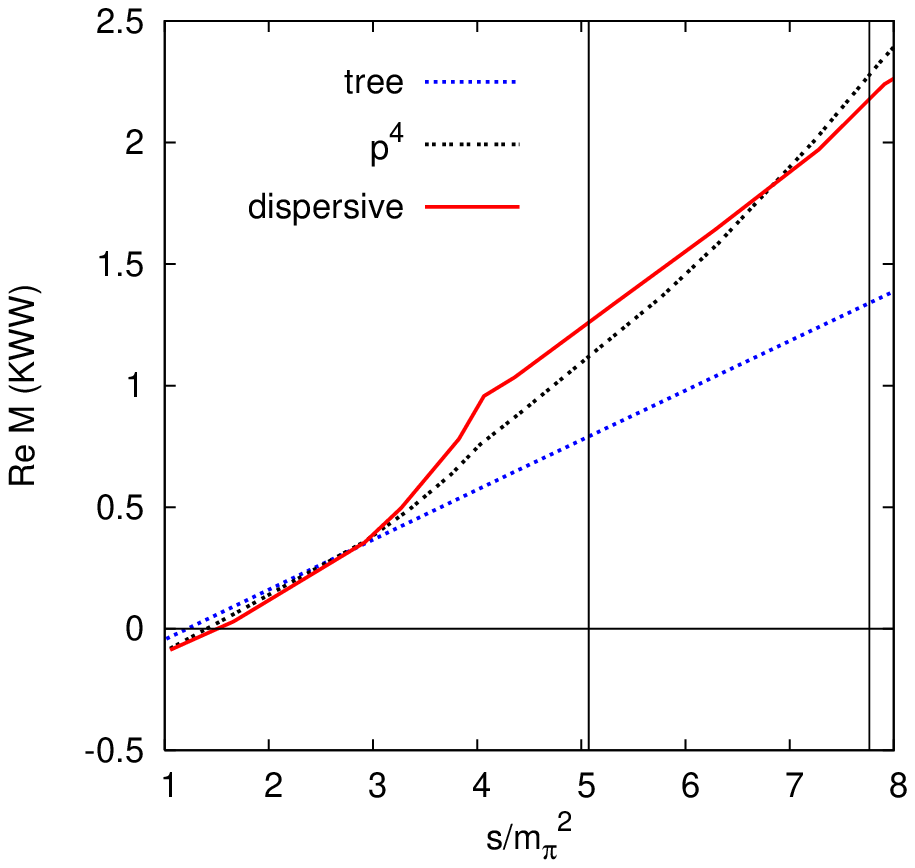}
\end{minipage}
\begin{minipage}{0.42\textwidth}
\includegraphics[width=\textwidth]{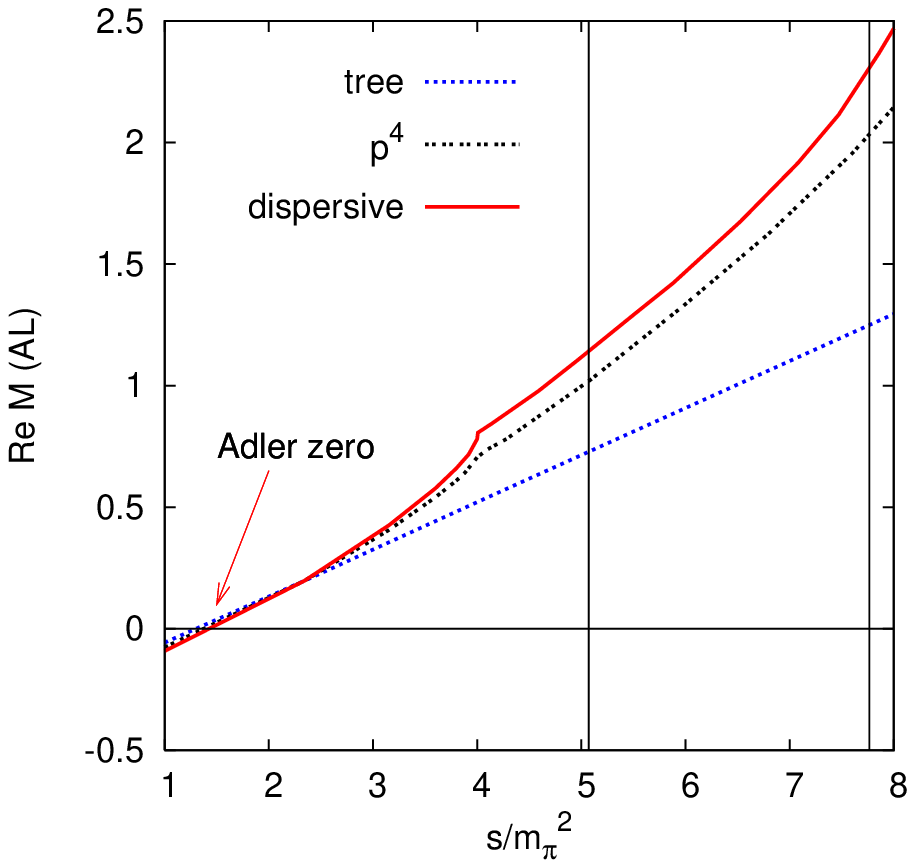}
\end{minipage}
\caption{
Left: Decay amplitude obtained by use of Khuri-Treiman
equations\cite{KWW} along the line $s=u$.
Right: Alternative dispersive analysis for the decay amplitude\cite{AL}.
Figs. from \cite{BijnensETA05}, adapted from \cite{KWW,AL}.
}
\label{figdispersive}
\end{figure}

In the limit of conserved isospin, no electromagnetism and
$m_u=m_d$, the $\eta$ is stable. Direct electromagnetic effects are 
small \cite{Sutherland1}.
The decay thus proceeds mainly through the quark-mass difference $m_u-m_d$.
The lowest order was done in \cite{orderp2x1},
order $p^4$ in
\cite{GL3} and recently the full order $p^6$ has been evaluated \cite{BG2}.
The momenta for the decay $\eta\to\pi^+\pi^-\pi^0$ we label as
$p_\eta$, $p_+$, $p_-$ and $p_0$ respectively and we introduce the
kinematical Mandelstam variables
$ 
s = (p_+ +p_-)^2\,, 
t = (p_+ +p_0)^2\,, 
u = (p_- +p_0)^2\,. 
$ 
These are linearly dependent,
$
s+t+u = m_{\pi^{o}}^2 + m_{\pi^{-}}^2 + m_{\pi^{+}}^2 + m_{\eta}^2
\equiv 3 s_0\,.  
$
The amplitudes for the charged, $A(s,t,u)$, and
neutral, $\overline{A}(s,t,u)$ are related
\ba
\overline{A}(s_1,s_2,s_3) &=& A(s_1,s_2,s_3)+A(s_2,s_3,s_1)+A(s_3,s_1,s_2)\, .
\label{defamplitude}
\ea
The relation in (\ref{defamplitude}) is only valid
to first order in $m_u-m_d$. The overall factor of $m_u-m_d$ can be put
in different quantities, two common choices are
\ba
A(s,t,u) = \frac{\sqrt{3}}{4R}M(s,t,u)
\quad&\mbox{or}&
\quad
A(s,t,u) =\frac{1}{Q^2} \frac{m_K^2}{m_\pi^2}(m_\pi^2-m_K^2)\,
\frac{ { {\cal M}(s,t,u)}}{3\sqrt{3}F_\pi^2}\,,
\label{defM}
\ea
with 
$R= (m_s-\hat m)/(m_d-m_u)$
or $Q^2 = R(m_s+m_d)/(2\hat m)$ pulled out.
The lowest order result is
\be
M(s,t,u)_{LO} = \left(({4}/{3})\,m_\pi^2-s\right)/F_\pi^2\,.
\label{LO}
\ee
The tree level determination of $R$ in terms of meson masses gives
with (\ref{LO}) a decay rate of 66~eV which should be compared with
the experimental results of 295$\pm$17~eV\cite{PDG06}.
In principle, since the decay rate is proportional to $1/R^2$ or $1/Q^4$,
this should allow for a precise determination of $R$ and $Q$. However,
the change required seems large. The order $p^4$ calculation
\cite{GL3} increased the predicted decay rate to 150~eV albeit with a
large error. About half of the enhancement in the amplitude came from
$\pi\pi$ rescattering and the other half from other effects like the
chiral logarithms\cite{GL3}. The rescattering effects have been
studied at higher orders using dispersive methods in \cite{KWW}
and \cite{AL}. Both calculations found an enhancement in the decay rate
to about 220~eV but differ in the way the Dalitz plot distributions
look. This can be seen in Fig.~\ref{figdispersive} where I show the 
real part of the amplitude as a function of $s$ along the line $s=u$.
The calculations use
different formalisms but make similar approximations, they mainly differ in
the determination of the subtraction constants.
That discrepancy and the facts that in $K_{\ell4}$ the dispersive estimate
\cite{BCG}
was about half the full ChPT calculation \cite{ABT3} and
at order $p^4$ the dispersive effect was about half of the correction for
$\eta\to3\pi$ makes it clear that also for this process a full order $p^6$
calculation was desirable.

\begin{figure}[h]
\vskip-0.5cm
\begin{minipage}{0.45\textwidth}
\includegraphics[width=\textwidth]{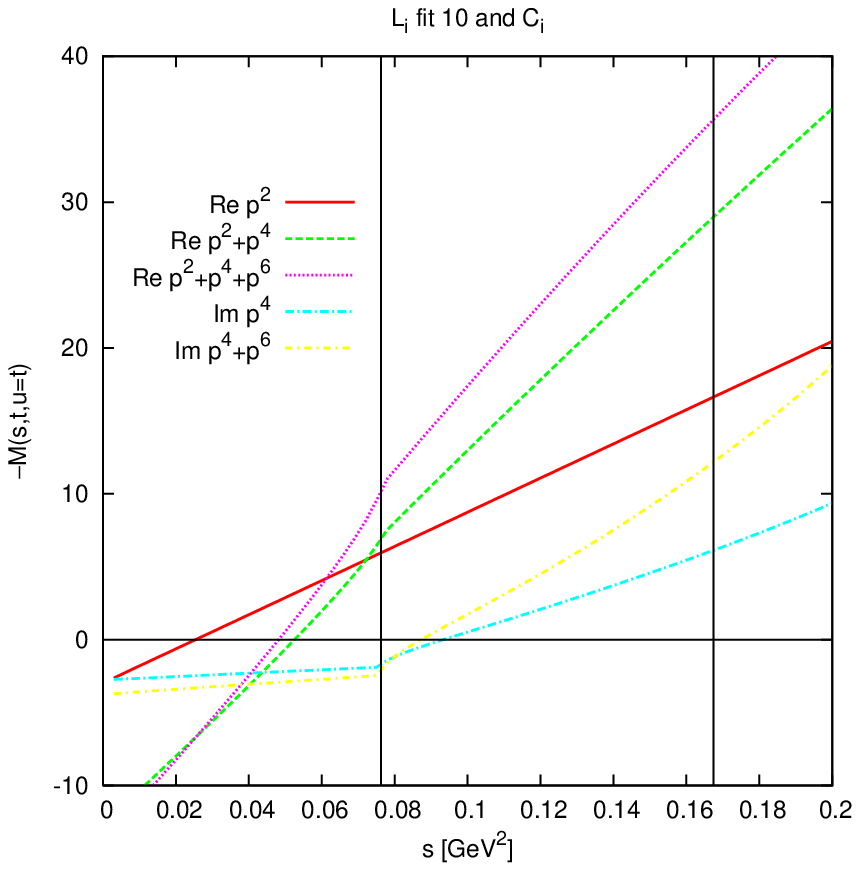}
\end{minipage}
\begin{minipage}{0.42\textwidth}
\rule{0cm}{0.5cm}\\
\includegraphics[width=\textwidth]{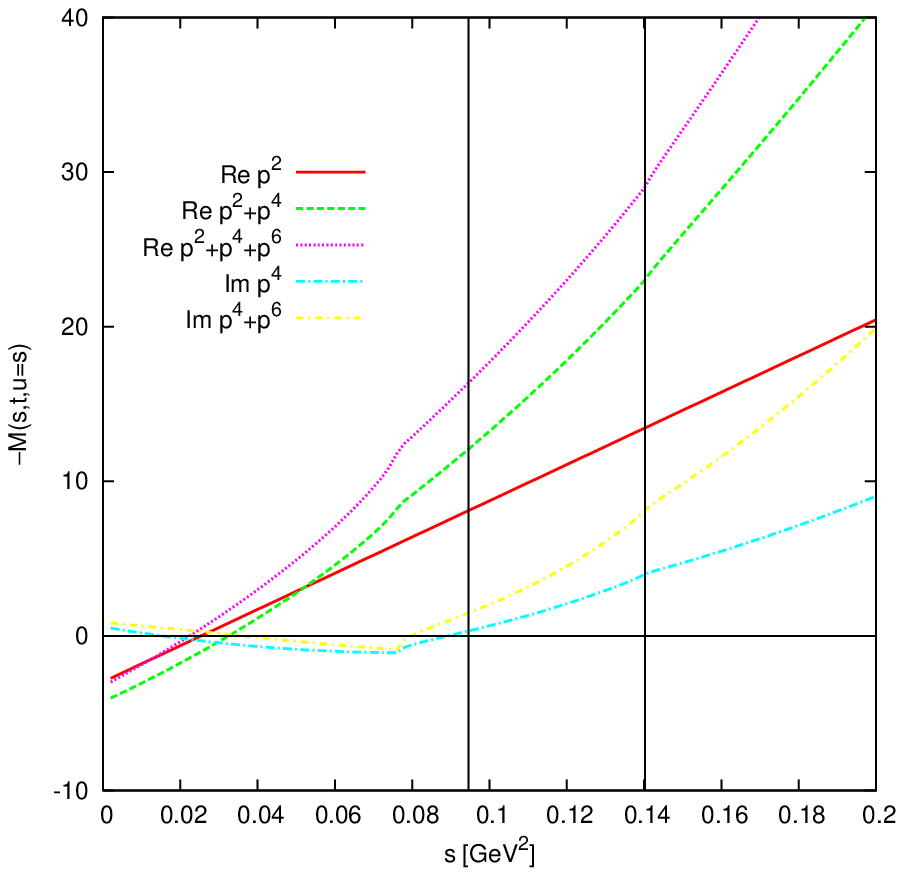}
\end{minipage}
\caption{Left: The amplitude $M(s,t,u)$ along the line $t=u$. The vertical
lines indicate the physical region.
Shown are the
real and imaginary parts with all parts summed up to the given order.
Right: Similar plot but along the line $s=u$.
Figs. from \cite{BG2}.}
\label{figMstu}
\end{figure}
The calculation \cite{BG2} generalized the methods
of \cite{ABT4} to deal with $\pi^0$-$\eta$ mixing. Here I show only results.
In Fig.~\ref{figMstu} I show the numerical result
for the amplitude along two lines in the Dalitz plot, $t=u$ and $s=u$.
The latter can be compared directly with the dispersive result of
Fig.~\ref{figdispersive}.
The correction found in \cite{BG2} at order $p^6$ is 20-30\% in amplitude,
larger in magnitude than the dispersive estimates \cite{KWW,AL} but
with a shape similar to \cite{AL}.

The Dalitz plot in $\eta\to3\pi$ is parameterized in terms of $x$ and $y$
defined in terms of the kinetic energies of the pions $T_i$ and
$Q_\eta=m_\eta-2m_{\pi^+}-m_{\pi^0}$ for the charged decay and $z$ defined in
terms of the pion energies $E_i$. The amplitudes are expanded in
$x = \sqrt3 \left(T_+-T_-\right)/Q_\eta$, $y= 3T_0/Q_\eta-1$,
$z = (2/3)\sum_{i=1,3}\left(3 E_i-m_\eta\right)^2/\left(m_\eta-3m_{\pi^0}\right)^2$, via
\ba
|M(s,t,u)|^2 &=& A_0^2\left(1+ay+by^2+dx^2+fy^3+\cdots\right)\,,
\quad
|\overline M(s,t,u)|^2 = \overline A_0^2
\left(1+2\alpha_2+\cdots\right)\,.
\ea
Recent experimental results for these parameters are shown in 
Tabs.~\ref{tabDalitzcharged} and \ref{tabDalitzneutral}.
There are discrepancies among the experiments but the latest
precision
measurements of $\alpha$ agree.
\begin{table}
\begin{center}
\small
\begin{tabular}{|c|ccc|}
\hline
Exp. & a & b & d  \\
\hline
\rule{0cm}{12pt}KLOE  & $-1.090$\small$\pm0.005^{+0.008}_{-0.019}$ &
 $0.124$\small$\pm0.006\pm0.010$ & $0.057$\small$\pm0.006^{+0.007}_{-0.016}$\\
Crystal Barrel & $-1.22\pm0.07$ & $0.22\pm0.11$ &
$0.06$\small$\pm0.04$ (input) \\
Layter {\it et al.} & $-1.08\pm0.014$ & $0.034\pm0.027$ &
$0.046\pm0.031$ \\
Gormley {\it et al} &$-1.17\pm0.02$ & $0.21\pm0.03$ 
& $0.06\pm0.04$ \\
\hline
\end{tabular}
\end{center}
\caption{Measurements of the Dalitz plot distributions in 
$\eta\to\pi^+\pi^-\pi^0$. Quoted in the order cited in \cite{etacharged}. 
The KLOE result $f$ is $f=0.14\pm0.01\pm0.02$.}
\label{tabDalitzcharged}
\end{table}
\begin{table}
\centerline{
\small
\begin{tabular}{|c|ccccc|}
\hline
               & $A_0^2$ & a & b & d & f \\
\hline
LO             & 120 & $-1.039$ & $0.270$ & $0.000$   & $0.000$ \\
NLO             & 314 & $-1.371$  & $0.452$ & $0.053$ & $0.027$\\
NLO ($L_i^r=0$) & 235 & $-1.263$& $0.407$ & $0.050$ & $0.015$\\
NNLO           & 538 &   $-1.271$ & $0.394$ & $0.055$ & $0.025$ \\
NNLO ($\mu=0.6$~GeV) & 543 & $-1.300$ & $0.415$ & $0.055$ & $0.024$\\         
NNLO ($\mu=0.9$~GeV) & 548 & $-1.241$ & $0.374$ & $0.054$ & $0.025$\\         
NNLO ($C_i^r = 0$)& 465 & $-1.297$  & $0.404 $ & $0.058$ & $0.032$ \\
NNLO ($L_i^r=C_i^r = 0$)& 251 & $-1.241$  & $0.424$ & $0.050$ & $0.007$ \\
\hline
\end{tabular}}
\caption{Theoretical estimate of the Dalitz plot distributions in 
$\eta\to\pi^+\pi^-\pi^0$.}
\label{tabDalitzcharged_theory}
\end{table}
The predictions from ChPT to order $p^6$ with the input parameters
as described earlier are given in Tabs.~\ref{tabDalitzcharged_theory}
and \ref{tabDalitzneutral_theory}. 
The different lines corresponds to variations on the input and the order of
ChPT. The lines labeled NNLO are the central results.
The agreement with experiment is not too good and clearly needs further study.
Especially puzzling is that $\alpha$ is consistently positive while
the dispersive calculations as well as \cite{Borasoy}
give a negative value.
The inequality $\alpha\le\left(d+b-a^2/4\right)/4$ derived in \cite{BG2}
shows that
$\alpha$ has rather large cancellations inherent in its prediction and that
the overestimate of $b$ is a likely cause of the wrong sign for
$\alpha$. The fairly large correction gives in the end
 larger values of $Q$ compared to those derived from the masses
\cite{BG2}.
\begin{table}
\begin{minipage}{0.56\textwidth}
\small
\begin{tabular}{|c|c|}
\hline
Exp. & $\alpha$\\
\hline
\rule{0cm}{12pt}
Crystal Ball (MAMI C) & $-0.032\pm0.003$\\
Crystal Ball (MAMI B) & $-0.032\pm0.002\pm0.002$\\
WASA/COSY  & $-0.027\pm0.008\pm0.005$\\
KLOE      & $-0.027\pm0.004^{+0.004}_{-0.006}$\\
Crystal Ball (BNL)  & $-0.031\pm0.004$ \\
WASA/CELSIUS   & $-0.026\pm0.010\pm0.010$ \\
Crystal Barrel & $-0.052\pm0.017\pm0.010$ \\
GAMS2000  & $-0.022\pm0.023$ \\
SND  & $-0.010 \pm 0.021 \pm 0.010$\\
\hline
\end{tabular}
\normalsize
\caption{Measurements of the Dalitz plot distribution in
$\eta\to\pi^0\pi^0\pi^0$. Quoted in the order cited in \cite{etaneutral}.} 
\label{tabDalitzneutral}
\end{minipage}
\begin{minipage}{0.43\textwidth}
\small
\begin{tabular}{|c|cc|}
\hline
    & $\overline A_0^2$ & $\alpha$\\
\hline
 LO &   1090 &  $ 0.000 $ \\
 NLO &  2810 &  $0.013$\\
NLO ($L_i^r=0$) & 2100 &   $ 0.016 $ \\
 NNLO &  4790 &  $ 0.013$ \\
NNLO ($C_i^r = 0$) & 4140 & $ 0.011$\\
NNLO ($L_i^r,C_i^r = 0$) & 2220 & $ 0.016$\\
\hline
dispersive \cite{KWW} & --- &\hskip-0.5cm $-(0.007$---$0.014)$\\
tree dispersive & --- & $-0.0065$\\
absolute dispersive& --- & $-0.007$\\
Borasoy \cite{Borasoy} & --- & $-0.031$\\
\hline
 error &  160 &  0.032 \\
\hline
\end{tabular}
\normalsize
\caption{Theoretical estimates of the Dalitz plot distribution in
$\eta\to\pi^0\pi^0\pi^0$. \cite{BG,BG2}}
\label{tabDalitzneutral_theory}
\end{minipage}
\end{table}

\section{Even more flavours at NNLO (or PQChPT)}

NLO Partially Quenched ChPT has been studied by many people and found
to be very useful, see \cite{Sharpelectures} and references therein.
The masses and decay constants are known to NNLO for almost all
possible mass combinations. Formulas were kept in terms of the quark-mass
expansion to avoid the proliferation
in physical masses appearing in this case. The three sea flavour masses
and decay constants are in \cite{BDL1,BL1} and the two sea
flavour results are in \cite{BL2}. Numerical programs are available
from the authors. The formulas are in the papers but can be downloaded
from \cite{website}.
PQChPT NNLO results for neutral masses are in \cite{BD1}.

\section{Renormalization Group}

In ChPT the renormalization group
is not quite as useful as in renormalizable theories but Weinberg
\cite{Weinberg0}
already showed that some predictions can indeed be made. In particular
leading logarithms at two loops can be had from only one-loop calculations.
This was used for obtaining the leading double logarithms in $\pi\pi$
scattering \cite{Colangelo1} and in general \cite{BCE0}.
The extension to all orders was proven in \cite{BC}.
A leading log to five loops was obtained in \cite{BF}
and recently recursion relations in the massless case were used to
get results to very high orders \cite{KPV1}.

The latter papers solved the practical problem of keeping track of all-order
Lagrangians using two observations. First, 
in the massless limit, tadpoles vanish,
and thus the number of external legs needed at any order does not increase.
Second, the main vertex needed is then the four meson vertex. Here they found
a useful general expressions using Legendre polynomials allowing them to do
all needed loop integrals and obtain a fairly simple iterative
algebraic recursion relation.

\section{Conclusions}

Modern ChPT is doing fine.
Two flavour ChPT is in good shape: it is now precision science in many ways.
For three flavour ChPT the corrections are larger and
there seem to be some problems, but many parameters, especially
in the scalar sector are rather uncertain and errors are very quantity
 dependent. Many partially quenched NNLO calculations
have been done with an eye on lattice calculations and their extrapolations.
A final comment is that new application areas continue to be found for
ChPT and EFT in general.

\section*{Acknowledgments}

This work is supported in part by the European Commission RTN network,
Contract MRTN-CT-2006-035482  (FLAVIAnet), 
European Community-Research Infrastructure
Integrating Activity
``Study of Strongly Interacting Matter'' (HadronPhysics2, Grant Agreement
n. 227431)
and the Swedish Research Council. I thank the organizers for a very pleasant
meeting.

\end{document}